\documentclass[conference]{IEEEtran}
\IEEEoverridecommandlockouts

\usepackage{filecontents}
\usepackage{cite}
\usepackage{amsmath,amssymb,mathtools,amsfonts}
\usepackage{algorithmic}
\usepackage{graphicx}
\usepackage{textcomp}
\usepackage{xcolor}
\usepackage[bookmarks=false]{hyperref}
\usepackage[tableposition=top,labelsep=period,font=footnotesize]{caption}

\usepackage{fancyhdr}
\usepackage{diagbox}
\usepackage{lipsum}
\usepackage{mathtools}
\usepackage{cuted}
\usepackage{pifont}
\usepackage{verbatim}
\usepackage{threeparttable}
\usepackage{ifthen}
\def\BibTeX{{\rm B\kern-.05em{\sc i\kern-.025em b}\kern-.08em
   T\kern-.1667em\lower.7ex\hbox{E}\kern-.125emX}}

\makeatletter
\def\ps@IEEEtitlepagestyle{%
    \def\@oddfoot{\mycopyrightnotice}%
    \def\@evenfoot{}%
}
\def\mycopyrightnotice{%
    \gdef\mycopyrightnotice{}
}

\begin{document}

\bstctlcite{IEEEexample:BSTcontrol}

\title{Optic-Net: A Novel Convolutional Neural Network for Diagnosis of Retinal Diseases from Optical Tomography Images\\
\thanks{ \textbf{Code Repository: }\textbf{\href{https://github.com/SharifAmit/OCT_Classification}{https://github.com/SharifAmit/OCT\textunderscore Classification}}}
}
\author{\IEEEauthorblockN{Sharif Amit Kamran\IEEEauthorrefmark{1}, Sourajit Saha\IEEEauthorrefmark{2}, Ali Shihab Sabbir\IEEEauthorrefmark{3} and Alireza Tavakkoli\IEEEauthorrefmark{4}}
\IEEEauthorblockA{\IEEEauthorrefmark{2}\IEEEauthorrefmark{3}
\textit{Center for Cognitive Skill Enhancement},
\textit{Independent University Bangladesh},
Dhaka, Bangladesh  \\
\IEEEauthorrefmark{1}\IEEEauthorrefmark{4}
\textit{University of Nevada, Reno},
NV, USA \\
skamran@nevada.unr.edu\IEEEauthorrefmark{1}, sourajit@iub.edu.bd\IEEEauthorrefmark{2}, asabbir@iub.edu.bd\IEEEauthorrefmark{3}, tavakkol@unr.edu\IEEEauthorrefmark{4}}
}

\maketitle

\begin{abstract}
Diagnosing different retinal diseases from Spectral Domain Optical Coherence Tomography (SD-OCT) images is a challenging task. Different automated approaches such as image processing, machine learning and deep learning algorithms have been used for early detection and diagnosis of retinal diseases. Unfortunately, these are prone to error and computational inefficiency, which requires further intervention from human experts. In this paper, we propose a novel convolution neural network architecture to successfully distinguish between different degeneration of retinal layers and their underlying causes. The proposed novel architecture outperforms other classification models while addressing the issue of gradient explosion. Our approach reaches near perfect accuracy of 99.8\% and 100\%  for two separately available Retinal SD-OCT data-set respectively. Additionally, our architecture predicts retinal diseases in real time while outperforming human diagnosticians. 
\end{abstract}


\section{Introduction}
Diabetes is a major health concern which affects up to 7.2\% of the population world-wide and the numbers could soon rise up to 600 million by the year 2040 \cite{centers2017national , yau2012global}. With its prevalence, one third of every diabetic patient develops Diabetic Retinopathy (DR). \cite{ting2016diabetic} This is a major cause for vision loss and affects nearly 2.8 \% of the population \cite{bourne2013causes}. Despite having effective vision tests for DR screening and early treatment in developed countries, avoiding erroneous results has always been a challenge for diagnosticians. On the other hand, DR has been often mistreated  in many developing and poorer economies, where access to trained ophthalmologist and eye-care machineries may be insufficient. So it's quite imminent to have an automated system which will help diagnose Diabetic Retinopathy and other related Retinal diseases with high precision and speed. This paper proposes a novel architecture based on convolutional neural network which can identify Diabetic Retionpathy, while being able to categorize multiple retinal diseases with near perfect accuracy. 

In ophthalmology a technique called Spectral Domain Optical Coherence Tomography (SD-OCT) is used for viewing the morphology of the retinal layers\cite{sri2014}. Moreover, depth-resolved tissue formation data encoded in the magnitude and delay of the back-scattered light by spectral analysis is also used to treat this diseases\cite{mlSD-OCT2017}. Though the image is retrieved through this process, the differential diagnosis is conducted by an ophthalmologist. Consequently, there will always be room for human error while performing the differential. Hence, an expert system is required to clearly distinguish between different retinal diseases with fewer mistakes.

One of the major reasons for misclassificaion is due to the stark similarity between Diabetic Retinopathy and other retinal diseases. They can be grouped by three major categories, i) Diabetic Macular Edema (DME) and Age-related degeneration of retinal layers (AMD), ii) Drusen, a condition where lipid or protein build-up occurs in the retinal layer and iii) Choroidal Neovascularization (CNV), a growth of new blood vessels in sub-retinal space. Diabetic Retinopathy and Age-related Macular Degeneration are the most likely cause of retinal diseases worldwide \cite{friedman2004prevalence}. While Drusen acts as an underlying cause that can trigger DR or AMD in a prolonged time-frame. On the other hand, Choroidal Neovascularization is an advanced stage of age-related macular degeneration that affects about 200,000 people worldwide every year \cite{ferrara2010vascular,wong2014global}.

Despite a decade of improvements to existing algorithms, identification of retinal diseases still produces erroneous results and requires expert intervention. To address this problem, we propose a novel architecture which not only identifies retinal diseases in real-time but also performs better than human experts for specific tasks. In the following sections, we elaborate our principal contributions and also provide a comparative analysis of different approaches. 

\begin{table*}[htp]
\centering
\caption{Comparison between different convolution operations used in the middle portion of residual unit.}
\begin{threeparttable}
\begin{tabular}{|c|c|c|c|}
\hline
 \begin{tabular}[c]{@{}c@{}}\textbf{Type of Convolution used in }\\ \textbf{ the Middle of Residual Unit}\end{tabular} & \begin{tabular}[c]{@{}c@{}}\textbf{ Approximate \# Parameters\textsuperscript{a}}\\ \textbf{( Without Counting Bias )}\end{tabular} & \begin{tabular}[c]{@{}c@{}}\textbf{Depletion Factor for Parameter, $\pmb{ \Phi_{p}}$}\\ \textbf{( Compared to Regular Convolution )} \end{tabular} &  \begin{tabular}[c]{@{}c@{}}\textbf{Test\textsuperscript{b}}\\ \textbf{Accuracy}\end{tabular} \\ \hline
\begin{tabular}[c]{@{}c@{}}Regular \\ Convolution\end{tabular} & $f^{2}\times D^{[i]}\times D^{[i-1]}$ = 36,864 & 100\% & 99.30\% \\ \hline
\begin{tabular}[c]{@{}c@{}}Atrous \\ Convolution\end{tabular} & $(f-1)^{2}\times D^{[i]}\times D^{[i-1]}$ = 16,384 & $(1-\frac{1}{f})^{2}$ = 44.9\% & 97.29\% \\ \hline
\begin{tabular}[c]{@{}c@{}}Separable \\ Convolution\end{tabular} & $(f^{2}+D^{[i]})\times D^{[i-1]}$ = 4,672 & $\frac{1}{f^{2}}+\frac{1}{D^{[i]}}$ = 12.5\% & 98.03\% \\ \hline
\begin{tabular}[c]{@{}c@{}}Atrous Separable \\ Convolution\end{tabular} & $((f-1)^{2}+D^{[i]})\times D^{[i-1]}$ = \textbf{4,352} & $\frac{1}{f^{2}}+(1-\frac{1}{f})^{2}\times \frac{1}{D^{[i]}}$ = \textbf{11.6\%} & 96.69\% \\ \hline
\begin{tabular}[c]{@{}c@{}c@{}}\textbf{Atrous Convolution}\\ \textbf{and Atrous Separable}\\ \textbf{Convolution Branched}\end{tabular} & $\frac{1}{2}((f-1)^{2}(1+\frac{1}{2}D^{[i]})+\frac{1}{2}D^{[i]})\times D^{[i-1]}$ = 5,248 & $\frac{1}{(2f)^{2}}+(1-\frac{1}{f})^{2}\times (\frac{1}{4}+\frac{1}{2D^{[i]}})$ = 14.4\% & \textbf{99.80\%} \\ \hline
\end{tabular}
\begin{tablenotes}\footnotesize
 \item[a]Here, kernel size, (f , f) = (3 , 3). Depth (\# kernels) in Residual unit's middle operation, $D^{[i]} = 64$ and first operation, $D^{[i-1]} = 64$.
 \item[b] The Test Accuracy reported in the table is obtained by training on OCT2017 \cite{kermany2018identifying} data-set, while the backbone network is Optic-Net 71.
\end{tablenotes}
\end{threeparttable}
\label{res_unit_tab}
\end{table*}

\section{Literature Review}
\subsection{Traditional Image Analysis}
The earliest approach to detect and classify retinal diseases from images included multiple image processing techniques followed by feature extraction and classification \cite{nguyen1997classification}. One such automated technique included finding abnormalities such as micro-aneurysms, haemorrhages, exudate and cotton wool-spot from Retinal Fundus images \cite{ege2000screening}. This approach uses a noise reduction algorithm and blurring to branch out the four-class problem to two cases of a  two-class  problem. From there on, background subtraction followed by shape estimation to extract important features is used. Finally, those features were used to classify each of the four abnormalities. Similarly, 
other such feature based technique was used for detecting Diabeitc Macular Edema and Choroidal Neovascularization. The  images were manipulated focused on five distinct parameters:   Retinal   Thickness,   augmentation  of Retinal Thickening, Macular volume, retinal morphology and vitreoretinal relationship \cite{panozzo2004diabetic}. Other approaches combined statistical classification with edge detection algorithms to detect sharp edges \cite{sanchez2004retinal}. Sanchez et al.'s \cite{sanchez2004retinal} algorithm achieved a sensitivity score of 79.6\% for classifying Diabeitc Retionpathy. Ege et al.'s \cite{ege2000screening} approach incorporating Mahalanobis classifier detected microaneurysms, haemorrhages, exudates, and cottonwool spots with a sensitivity of 69, 83, 99, and 80\%, respectively. It's quite evident that each of these techniques shown slight improvements, but in terms of precision it didn't achieve desired results. 

\subsection{Segmentation based approaches}
The most notable way to identify a patient having Diabetic Macular Edema is the enlargement of macular density in retinal layer \cite{costa2006retinal,mlSD-OCT2017}. Many approaches have been proposed and implemented that involves segmentation of retinal layers. Further identification of likely causes are also performed for build-up of liquids in the sub-retinal space \cite{meindertniemeijer20123d,lang2013retinal,mishra2009intra}. In \cite{quellec2010three,lee2010segmentation}, the authors proposed the idea of segmenting the intra-retinal layers in ten parts and then extracted the texture and depth information from each layer. Subsequently, any aberrant retinal features are detected by classifying the dissimilarity between healthy retinas and the diseased ones. Niemeijer et al. \cite{meindertniemeijer20123d} introduced a technique for 3D segmentation of regions containing fluid in OCT images using a graph-based implementation. A graph-cut algorithm is applied to get the final predictions from the information initially retrieved from layer-based segmentation of fluid regions. Even though implementation based on a previous segmentation of retinal layers have reported high scoring prediction results, the initial step is reportedly troublesome and erroneous \cite{ghorbel2011automated,kafieh2013review}. As reported in \cite{lee2013fully}, retinal thickness measurements obtained by different systems has stark dissimilarity. Therefore, it is not quite effective to compare between different retinal depth information retrieved by separate machines. Enforcing the fact that segmentation based approaches weren't effective as a universal retinal disease recognition system.

\subsection{Machine Learning and Deep Learning techniques}
 Lately, a combination of machine learning and deep learning architectures has become a go to, for achieving state-of-the-art accuracy for recognizing various retinal diseases \cite{lemaitre2016classification,lee2017deep,treder2018automated}. Awais et al. combined VGG16 \cite{simonyan2014very} with KNN and Random forest classifier (100 trees) to create a deep classification architecture for differentiating between Normal Retina and Diabetic Macular Edema. On the other hand, Lee et al. used a standalone VGG16 architecture with a binary output to detect Age-related Macular Edema (AMD) \cite{lee2017deep}. Although this techniques exploit automatic feature learning from large array of images, the architecture itself isn't efficient in terms of speed and memory usage. On the contrary, transfer learning methods depend on weeks of training on millions of images and are not idle for finding stark differences between Retinal diseases. To help alleviate from all of these challenges an architecture is necessary which is specially catered for identifying retinal deceases with high precision, speed, and low memory usage. 
 
\begin{figure*}[htp]
\centering
\includegraphics[width=16cm,height=10.1cm]{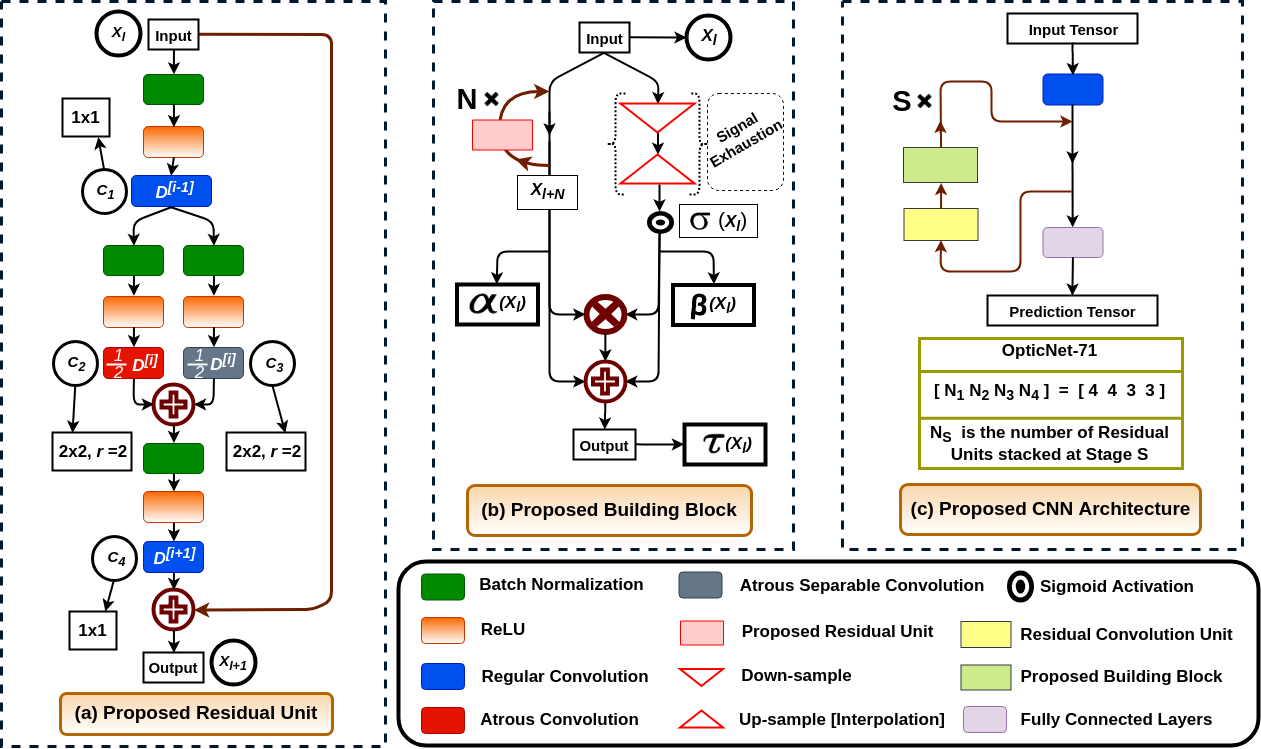}
\caption{Illustration of the Building Blocks of our proposed CNN \textbf{[ OpticNet-71 ]}. Only the very first convoluiton (\(7 \times 7\)) layer in the CNN and the very last convolution (\(1 \times 1\)) layer in Stage[2,3,4]: Residual Convolutional Unit uses stride 2, while all other convolution operations use stride 1.}
\label{full_cnn_fig}
\end{figure*}
 
\subsection{Our Contributions}
In this work, we propose a novel convolutional neural network which specializes in identifying retinal diseases with near perfect precision. Moreover, through this architecture we are proposing (a) a new residual unit subsuming Atrous Separable Convolution, (b) a novel building block and (c) a mechanism to prevent gradient degradation. The proposed network outperforms other architectures with respect to the  number of parameters, accuracy, and memory size. Our proposed architecture is trained from scratch and bench-marked on two publicly available data-sets: OCT2017 \cite{kermany2018identifying}, Srinivasan2014 \cite{sri2014} data-sets. Henceforth, it doesn't require any pre-trained weights, reducing the training and deployment time of the model by many folds. We believe with the deployment of this model, the rapid identification and treatment can be carried out with near perfect certainty. Additionally, it will aid the ophthalmologist to get a second expert opinion for their differential diagnosis. 

\section{Proposed Methodology}
Fig.~\ref{full_cnn_fig} illustrates the Deep Convolutional Neural Network (CNN) architecture we propose for the classification of retinal diseases from Optical Coherence Tomography (OCT) images. In Fig.~\ref{full_cnn_fig}(a) we delineate how the proposed Residual Learning Unit improves feature learning capabilities while discussing the techniques we adopt to reduce computational complexity for such performance enhancement. While, Fig.~\ref{full_cnn_fig}(b) depicts the proposed mechanism to handle gradient degradation, Fig.~\ref{full_cnn_fig}(c) narrates the entire CNN architecture. We discuss the constituent segments of our CNN architecture, called Optic-Net, over the following subsections.

\subsection{Proposed Residual Learning Mechanism}
Historically, Residual Units \cite{he2016deep,he2016identity} used in Deep Residual Convolutional Neural Networks (CNN), process the incoming input through three convolution operations while adding the incoming input with the processed output. These three convolutional operations are (1$\times$1), (3$\times$3) and (1$\times$1) convolutions. Therefore, replacing the (3$\times$3) convolution in the middle with other types of convolutional operations can potentially change the learning behaviour, computational complexity and eventually prediction performance.

We experimented with different convolution operations as replacement for the (3$\times$3) middle convolution and observed which choice contributes the most to reduce the number of parameters, ergo computational complexity, as depicted in Table \ref{res_unit_tab}. Furthermore, in Table \ref{res_unit_tab}, we use a depletion factor for parameters, $\Phi_{p}$ which is a ratio of number of parameters in the replaced convolution and regular convolution expressed in percent. The first four rows of Table \ref{res_unit_tab} indicates that using Atrous Separable Convolution is the most computationally effective method. However, our experiment shows that this does not lead to the best prediction performance, which we demonstrated in Table \ref{res_unit_tab} as well. 

In this work however, we replace the middle (3$\times$3) convolution operation with two different operations running in parallel as detailed in Fig.~\ref{full_cnn_fig}(a). Whereas, a conventional residual unit uses $D^{[i]}$ number of channels for the middle convolution, we use $\frac{1}{2}D^{[i]}$ number of channels for each of the newly replaced operations to prevent any surge in parameter. In the proposed branching operation we use a (2$\times$2) Atrous convolution ($C_{2}$) with dilation rate, $r$ = 2 to get a (3$\times$3) receptive field in the left branch while in the right branch we use a (2$\times$2) Atrous separable convolution ($C_{3}$) with dilation rate, $r$ = 2 to get a (3$\times$3) receptive field. Sequentially, the results are then added together. Furthermore, separable convolution \cite{sifre2014rigid} disentangles the spatial and depth-wise feature maps separately while Atrous convolutions inspect both spatial and depth channels together. We hypothesize that adding two such feature maps that are learned very differently shall help trigger more robust and subtle features.

\begin{equation}
\begin{split}
X_{l+1} &= X_{l} + \big[(X_{l} \circledast C_{1} \circledast C_{2}) + (X_{l} \circledast C_{1} \circledast C_{3})\big] \circledast C_{4} \\&= X_{l} + \hat{F}(X_{l},W_{l}) \label{res_unit_eqn}
\end{split}
\end{equation}

\begin{figure}[t]
\centering
\includegraphics[width=8.8cm,height=8cm]{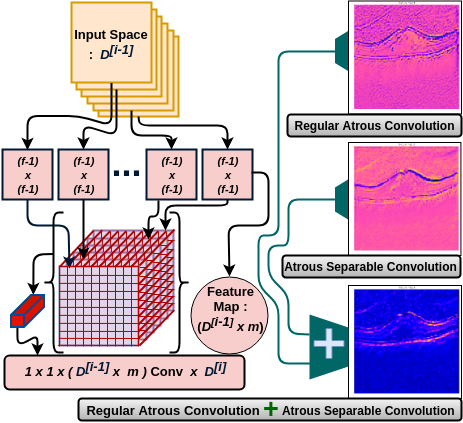}
\caption{Atrous Separable Convolution. Exploiting $(f-1) \times (f-1)$ convolutions in stead of $f \times f$ that yields more fine grained and coarse features with better depth resolution compared to regular Atrous Convolution.}
\label{atr_sep_fig}
\end{figure}

Fig.~\ref{atr_sep_fig} shows how adding Atrous and Atrous separable feature maps help disentangle the input image space with more depth information instead of activating only the predominant edges. Moreover, the last row of Table \ref{res_unit_tab} confirms that adopting this strategy still reduces the computational complexity by a reasonable margin, while improving inference accuracy. Equation \eqref{res_unit_eqn} further clarifies how input signals $X_{l}$ travel through the proposed residual unit shown in Fig.~\ref{full_cnn_fig}(a), where $\circledast$ refers to convolution operation.

\subsection{Proposed Building Block and Signal Propagation}
In this section we discuss the proposed building block as a constituent part of Optic-Net. As shown in Fig.~\ref{full_cnn_fig}(b) we split the input signal ($X_{l}$) into two branches - (1) Stack of Residual Units, (2) Signal Exhaustion. Later in this section we explain how we connect these two branches to propagate signals further in the network.

\begin{equation}
\alpha(X_{l}) = X_{l+N} = X_{l} + \sum_{i=l}^{N} \hat{F}(X_{i},W_{i})  \label{stacked_res_unit_eqn}
\end{equation}

\subsubsection{Stack of Residual Units}
In order to initiate a novel learning chain for propagating signals through stacking several of the proposed residual units linearly, we suggest to combine global residual effects enhanced by pre-activation residual units \cite{he2016identity} and our proposed set of convolution operations (Fig.~\ref{full_cnn_fig}(a)). As shown in \eqref{res_unit_eqn}, $\hat{F}(X_{l},W_{l})$ denotes all the proposed set of convolution operations inside a residual unit for input $X_{l}$. We sequentially stack these residual units N times over $X_{l}$ which is input to our proposed building block, as narrated in Fig.~\ref{full_cnn_fig}(b). Equation \eqref{stacked_res_unit_eqn} illustrates the state of output signal denoted by $X_{l+N}$ which is processed through a stack of residual units of length $N$. For the sake of further demonstration we denote $X_{l+N}$ as $\alpha(X_{l})$.

\subsubsection{Signal Exhaustion}
In the proposed building block, we propagate the input signal $X_{l}$ through an Max-pooling layer to achieve spatial down-sampling which we then up-sample through Bi-linear interpolation. Since the down-sampling module only forwards the strongest activations, the interpolated reconstruction makes a dense spatial volume from the down-sampled representation - intrinsically exhausting the incoming signal $X_{l}$. As detailed in Fig.~\ref{full_cnn_fig}(b), we sequentially pass the exhausted signal space through sigmoid activation, $\sigma(X_{l})=1/(1+e^{-X_{l}})$. Recent research \cite{wang2017residual} has shown how auto-encoding with residual skip connections [\(P_{encoder}(input|code)\) \(\mapsto\) \(P_{decoder}(code|input) + input\)] improve attention oriented classification performance. However unlike auto-encoders, max-pooling and Bi-linear interpolation functions are not enabled with learning mechanism. In Optic-Net, we capacitate the CNN to activate spikes from a exhausted signal space because we use it as a mechanism to avert gradient degradation. For the sake of further demonstration we denote the exhausted signal activation module, $\sigma(X_{l})$ as $\beta(X_{l})$.

\begin{table}[b]
\centering
\caption{Architectural Specifications for Opticnet-71 and Layer-wise Analysis for Number of Feature Maps in Comparison with Resnet50-v1 \cite{he2016deep}.}
\begin{tabular}{|c|c|c|}
\hline
\textbf{Layer Name}&\textbf{ResNet50 V1\cite{he2016deep}}&\textbf{OpticNet71 [Ours]}\\ \hline
Conv \(7\times7\)&\([64]\times1\)&\([64]\times1\)\\ \hline
Stage1: Res Conv&\([64,64,256]\times1\)&\([64,64,256]\times1\)\\ \hline
Stage1: Res Unit&\([64,64,256]\times2\)&\([32,32,32,256]\times4\)\\ \hline
Stage2: Res Conv&\([128,128,512]\times1\)&\([128,128,512]\times1\)\\ \hline
Stage2: Res Unit&\([128,128,512]\times3\)&\([64,64,64,512]\times4\)\\ \hline
Stage3: Res Conv&\([256,256,1024]\times1\)&\([256,256,1024]\times1\)\\ \hline
Stage3: Res Unit&\([256,256,1024]\times5\)&\([128,128,128,1024]\times3\)\\ \hline
Stage4: Res Conv&\([512,512,2048]\times1\)&\([512,512,2048]\times1\)\\ \hline
Stage4: Res Unit&\([512,512,2048]\times2\)&\([256,256,256,2048]\times3\)\\ \hline
Global Avg Pool& 2048 & 2048\\ \hline
Dense Layer 1&K (Classes) & 256\\ \hline
Dense Layer 2& -- & K (Classes)\\ \hline
Parameters& 25.64 Million & \textbf{12.50 Million}\\ \hline
Required FLOPs& 3.8 $\times$ $10^{9}$ & \textbf{2.5} $\pmb{\times}$ $\pmb{10^{7}}$ \\ \hline
CNN Memory& 98.20 MB & \textbf{48.80 MB} \\ \hline
\end{tabular}
\label{cnn_table}
\end{table}

\begin{equation}
\tau(X_{l}) = \alpha(X_{l})+\beta(X_{l})+\alpha(X_{l})\times\beta(X_{l}))  \label{sig_prop_eqn}
\end{equation}

\begin{figure*}[htp]
\begin{equation}
\begin{split}
\frac{\partial\tau(X_{l})}{\partial X_{l}} &= \bigg[ \bigg( 1 + \sigma(X_{l}) \bigg) \times \bigg( 1 + \frac{\partial}{\partial X_{l}}\sum_{i=l}^{N} \hat{F}(X_{i},W_{i}) \bigg) \bigg] + \bigg[ \bigg( 1 + X_{l} + \sum_{i=l}^{N} \hat{F}(X_{i},W_{i}) \bigg) \times \bigg( \sigma(X_{l}) \bigg) \times \bigg( 1 - \sigma(X_{l}) \bigg) \bigg] \\&= \bigg[ \bigg( 1 + \frac{\sigma^{\prime}(X_{l})}{1-\sigma(X_{l})} \bigg) \times \bigg( 1 + \frac{\partial}{\partial X_{l}}\sum_{i=l}^{N} \hat{F}(X_{i},W_{i}) \bigg) \bigg] + \bigg[ \bigg( 1 + X_{l+N} \bigg) \times \bigg( \sigma^{\prime}(X_{l}) \bigg) \bigg] \label{block_grad_eqn}
\end{split}
\end{equation}
\end{figure*}

\subsubsection{Signal Propagation} \label{sig_prop}
As shown if Fig.~\ref{full_cnn_fig}(b), we process the residual signal, $\alpha(X_{l})$ and exhausted signal, $\beta(X_{l})$ following \eqref{sig_prop_eqn} and we denote the output signal propagated from the proposed building block as $\tau(X_{l})$. Our hypothesis behind such design is that, whenever one of the branch falls prey to gradient degradation from a mini-batch the other branch manages to propagate signals unaffected by the mini-batch with amplified absolute gradient. To validate our hypothesis \eqref{sig_prop_eqn} shows that, \(\tau(X_{l})\approx\alpha(X_{l}), \forall\beta(X_{l})\approx 0\) and \(\tau(X_{l})\approx\beta(X_{l}), \forall\alpha(X_{l})\approx 0\) illustrating how the unaffected branch survives the degradation in the affected branch. However, when none of the branch gets affected by gradient amplification the multiplication ($\alpha(X_{l}) \times \beta(X_{l})$) balances out the increase in signal propagation due to both branch's addition. Equation \eqref{block_grad_eqn} delineates the gradient of building block output $\tau(X_{l})$ with respect to  building block input $X_{l}$ calculated during back-propagation for optimization.

\subsection{CNN Architecture and The Optimization Chain}
Fig. \ref{full_cnn_fig}(c) portrays the entire CNN architecture with all the building blocks and constituent components joined together. First, the input batch (224\(\times\)224\(\times\)3) is propagated through an 7\(\times\)7 Conv with stride 2 that follows batch-normalization and ReLU  activation. Then we propagate the signals via a Residual Convolution Unit (same as the unit used in \cite{he2016identity}) which is then followed by our proposed building block. We propagate the signals through this [Residual Convolution Unit \(\mapsto\) Building Block] procedure for S = 4 times, as we call them stage 1, 2, 3 and 4 respectively. Then global average pooling is applied to the signals which passes through two more Fully Connected(FC) layers for the loss function which is denoted by \(\Delta\xi\).

In Table \ref{cnn_table}, we show the number of feature maps (Layer Depth) we use for each layer in the network. The output shape of the input tensor after four consecutive stages are (112\(\times\)112\(\times\)256), (56\(\times\)56\(\times\)512), (28\(\times\)28\(\times\)1024) and (14\(\times\)14\(\times\)2048) respectively. Moreover, $FC_{1}$ $\in \mathbb{R}^{2048 \times 256}$ and $FC_{2}$ $\in \mathbb{R}^{256 \times K}$ ,where K = number of classes.

\begin{equation}
\frac{\partial\xi}{\partial X_{l}} = \frac{\partial\xi}{\partial\tau(X_{l}^{Stage})} \times \prod_{j=1}^{Stage}\bigg[\frac{\partial\tau(X_{l}^{j})}{\partial X_{l}^{j}}\bigg] \label{full_cnn_eqn}
\end{equation}

Equation \eqref{full_cnn_eqn} represents the gradient calculated for the entire network chain distributed over stages of optimization. As \eqref{block_grad_eqn} suggests, the term $(1 + \sigma(X_{l}))$ - in comparison with \cite{he2016identity}) - works as an extra layer of protection to prevent possible gradient explosion caused by the stacked residual units by multiplying non-zero activations with the residual unit's gradients. Moreover, the term $(1 + X_{l+N})$ indicates that the optimization chain still has access to signals from much earlier in the network and to prevent unwanted spikes in activations the term $\sigma^{\prime}(X_{l})$ can still mitigate gradient expansion which can potentially jeopardize learning otherwise.

\begin{table}[b]
\centering
\caption{ Penalty Weights Proposed for Oct2017 \cite{kermany2018identifying} }
\begin{threeparttable}

\begin{tabular}{|c|c|c|c|c|}
\hline
       & Normal & Drusen & CNV\textsuperscript{1} & DME\textsuperscript{2} \\ \hline
Normal & 0      & 1      & 1   & 1   \\ \hline
Drusen & 1      & 0      & 1   & 1   \\ \hline
CNV\textsuperscript{1}    & 4      & 2      & 0   & 1   \\ \hline
DME\textsuperscript{2}    & 4      & 2      & 1   & 0   \\ \hline
\end{tabular}
\begin{tablenotes}
     \item[1] CNV : Chorodial Neovascularization
     \item[2] DME : Diabetic Macular Edema
   \end{tablenotes}
\end{threeparttable}
\label{table2}
\end{table}

\begin{figure}[b]
\centering
\includegraphics[width=8cm,height=4.75cm]{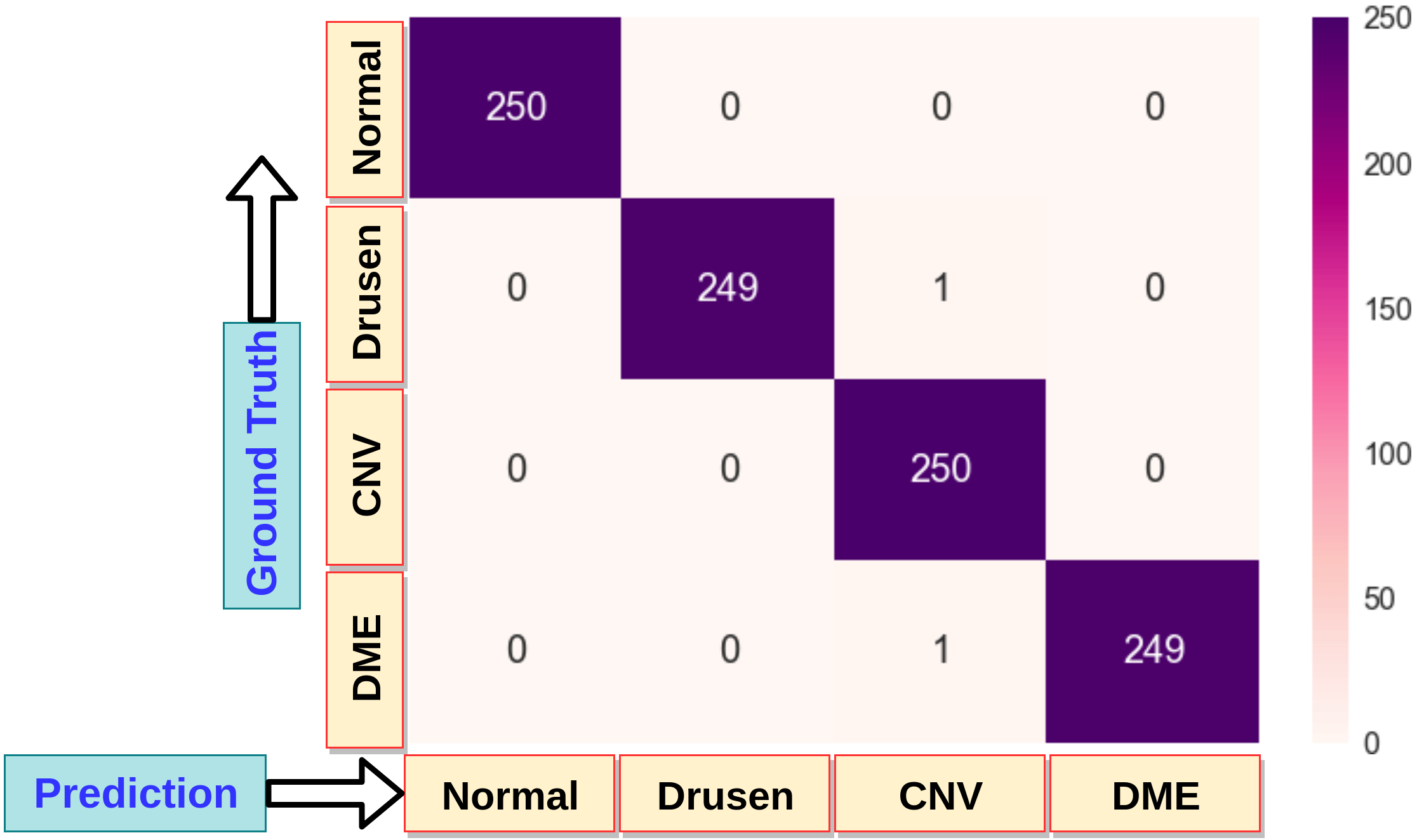}
\caption{Confusion matrix generated by OpticNet-71 for OCT2017\cite{kermany2018identifying} data-set.}
\label{fig3}
\end{figure}

\section{Experiments}
\subsection{Specifications of Data-sets and Pre-processing Techniques}
We benchmark our model against two distinct data-sets (different scale, sample space, etc.). The first data-set aims at correctly recognizing and differentiating between four distinct retinal states provided by the OCT2017 \cite{kermany2018identifying} data-set. Where, the stages are normal healthy retina, Drusen, Choroidal Neovascularization (CNV) and Diabetic Macular Edema (DME). OCT2017 \cite{kermany2018identifying} data-set contains 84,484 images (provided as high quality TIFF format with 3 non-RGB color channels). We split them into 83,484 train-set and 1000 test-set. The second data-set - Srinivasan2014 \cite{sri2014} - consists of three classes and aims at classifying normal healthy specimen of retina, Age-Related Macular Degeneration (AMD) and Diabetic Macular Edema (DME). Srinivasan2014 \cite{sri2014} data-set consists of 3,231 image samples that we split into 2,916 train-set, 315 test-set. We resize images from both data-sets to $224\times224\times3$ for both training and testing. For both the data-set we do 10-fold cross-validation on the training set and find the best models.

\subsection{Performance Metrics}
We calculated four standard metrics to evaluate our CNN model on both data-sets : Accuracy \eqref{8}, Sensitivity \eqref{9}, Specificity \eqref{10} and a Special Weighted Error \eqref{11} from \cite{kermany2018identifying}. Where N is the number of image samples and K is the number of classes. Here TP, FP, FN and TN denotes True Positive, False Positive, False Negative and True Negative respectively. We report True Positive Rate (TPR) or Sensitivity \eqref{8} and True Negative Rate (TNR) or Specificity \eqref{9} for the both the data-sets \cite{sri2014,kermany2018identifying}. For this, we calculate the TPR and TNR for individual classes then sum all the values and then divide that by the number of classes (K).
\begin{equation}
\textbf{Accuracy = }\frac{1}{N}\sum TP \label{8}
\end{equation}
\begin{equation}
\textbf{Sensitivity = }\frac{1}{K}\sum \frac{TP}{TP + FN} \label{9}
\end{equation}
\begin{equation}
\textbf{Specificity = }\frac{1}{K}\sum \frac{TN}{TN + FP} \label{10}
\end{equation}
\begin{equation}
\textbf{Weighted Error = }\frac{1}{N}\sum_{i,j \in K} W_{ij}\cdot X_{ij} \label{11}
\end{equation}
As reported in \cite{kermany2018identifying}, the penalty points for incorrect categorization of a retinal disease can be arbitrary. Table \ref{table2} shows the penalty weight values for misidentifying a category set by \cite{kermany2018identifying} which is only specific to OCT2017 \cite{kermany2018identifying} data-set. To calculate Weighted Error \eqref{11}, we apply element-wise multiplication on the confusion matrix generated by specific model (Fig. \ref{fig3} represents the confusion matrix generated by OpticNet-71 on OCT2017 \cite{kermany2018identifying} data-set) and the weight matrix in Table \ref{table2} and then take an average over the number of samples. Here, the penalty weight values from Table \ref{table2} is denoted by W and the model's prediction (confusion matrix) is denoted by X where i,j denotes the rows and columns of the confusion matrix.
\begin{figure}[b]
\centering
\includegraphics[width=8.7cm,height=5.98cm]{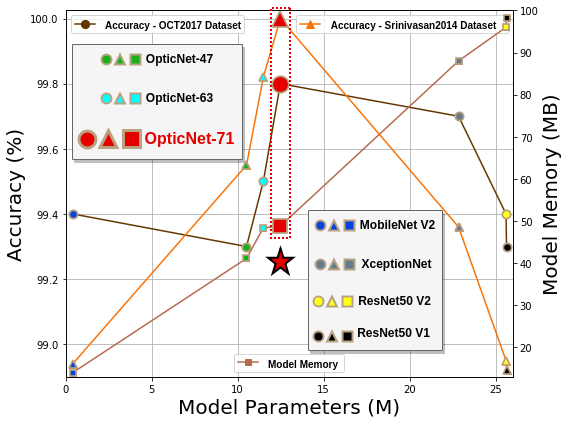}
\caption{Test accuracy (\%), CNN memory (Mega-Bytes) and model parameters (Millions) on OCT2017 \cite{kermany2018identifying} data-set and Srinivasan2014 \cite{sri2014} data-set.}
\label{fig4}
\end{figure}
\begin{figure*}[t]
\centering
\includegraphics[width=18cm,height=8.3cm]{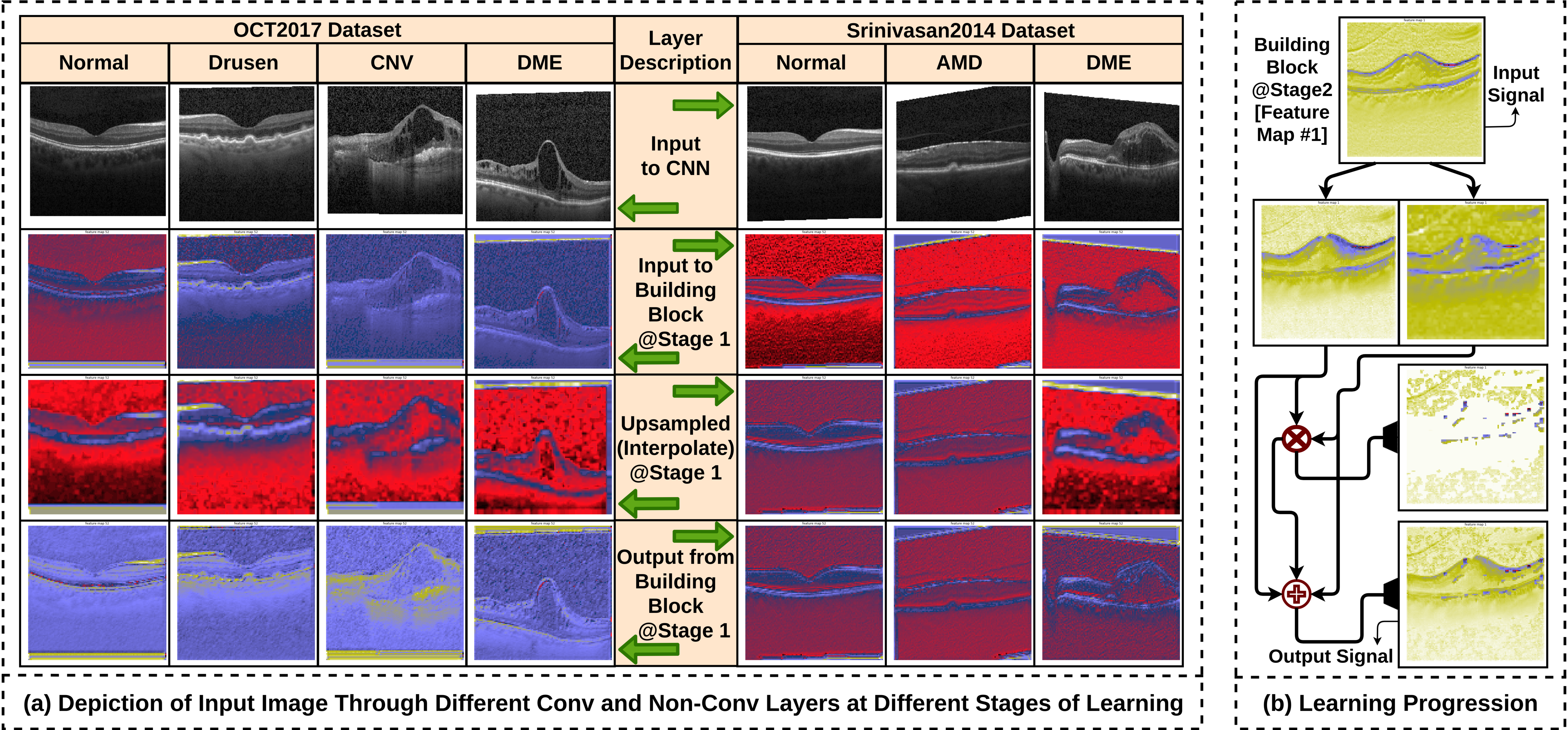}
\caption{(a) Visualizing input images from each class through different layers of Optic-Net 71. As shown, the feature maps at the end of each building block learns more fine grained features by focusing sometimes on the same shapes - rather in different regions of the image - learning to decide what features lead the image to the ground truth. (b) The learning progression however, shows how exhausting the signal propagated with residual activation learns to detect more thin edges - delving further into the Macular region to learn anomalies. While using the signal exhaustion mechanism sometimes, important features can be lost during training. Our experiments show, by using more of these building blocks we can reduce that risk of feature loss and improve overall optimization for Optic-Net 71.}
\label{fig5}
\end{figure*}

\subsection{Training OpticNet-71 and Obtained Results}
\subsubsection{OCT2017 Data-set}
In Table \ref{table3}, we report a comprehensive study for OCT2017 \cite{kermany2018identifying} data-set evaluated through testing standards such as Test Accuracy, Sensitivity, Specificity and Weighted Error. OpticNet-71 scores the highest Test Accuracy ($99.80\%$) among other existing solutions, with a Sensitivity and Specificity of $99.80\%$ and $99.93\%$. Furthermore, the Weighted Error is reported to be a mere $0.20\%$ which can be visualized in Fig.~\ref{fig3} as our architecture misidentifies one Drusen and one DME sample as CNV. However, the penalty weight is only 1 for each of the mis-classification as we report in Table \ref{table2}. Sequentially, with our proposed OpticNet-71 we obtain state-of-the-art results on OCT2017 \cite{kermany2018identifying} data-set across all four performance metrics, while significantly surpassing human benchmarks as mentioned in Table \ref{table3}.

\begin{table}[tp]
\centering
\caption{Results on Oct2017 \cite{kermany2018identifying} Data-set.}
\begin{tabular}{|c|c|c|c|c|}
\hline
Architectures                                                             & \begin{tabular}[c]{@{}c@{}}Test\\ Accuracy\end{tabular} & Sensitivity & Specificity & \begin{tabular}[c]{@{}c@{}}Weighted\\ Error\end{tabular} \\ \hline
\begin{tabular}[c]{@{}c@{}}InceptionV3\\ (limited)\end{tabular}           & 93.40                                                   & 96.60       & 94.00       & 12.70                                                    \\ \hline
\begin{tabular}[c]{@{}c@{}}Human\\ Expert 2\cite{kermany2018identifying}  \end{tabular}                  & 92.10                                                   & 99.39       & 94.03       & 10.50                                                    \\ \hline
InceptionV3\cite{kermany2018identifying}                                                                 & 96.60                                                   & 97.80       & 97.40       & 6.60                                                     \\ \hline
ResNet50-v1 \cite{he2016deep}                                                               & 99.30                                                   & 99.30       & 99.76       & 1.00                                                     \\ \hline
MobileNet-v2\cite{sandler2018mobilenetv2}                                                              & 99.40                                                   & 99.40       & 99.80       & 0.60                                                     \\ \hline
\begin{tabular}[c]{@{}c@{}}Human \\ Expert 5 \cite{kermany2018identifying}\end{tabular}                 & 99.70                                                   & 99.70       & 99.90       & 0.40                                                     \\ \hline
Xception \cite{chollet2017xception}                                                                    & 99.70                                                   & 99.70       & 99.90       & 0.30                                                     \\ \hline
\textbf{\begin{tabular}[c]{@{}c@{}}OpticNet-71\\ {[}Ours{]}\end{tabular}} & 99.80                                                   & 99.80       & 99.93       & 0.20                                                     \\ \hline
\end{tabular}
\label{table3}
\end{table}

\subsubsection{Srinivasan2014 Data-set}
We benchmark OpticNet-71 against other methods in Table \ref{table4} while evaluating Srinivasan2014\cite{sri2014} data-set through three metrics: Accuracy, Sensitivity and Specificity. Among the mentioned solutions in Table \ref{table4} Lee et al. \cite{lee2017deep} uses modified VGG-16, Awais et al. \cite{awais2017classification} uses VGG architecture with KNN in final layer and Karri et al. \cite{karri2017transfer} uses GoogleNet while they all use weights from transfer learning on ImageNet \cite{russakovsky2015imagenet}. As shown in Table \ref{table4}, OpticNet-71 achieves state-of-the-art result by scoring $100\%$ Accuracy, Sensitivity and Specificity.

Furthermore, we train ResNet50-v1\cite{he2016deep}, ResNet50-v2\cite{he2016identity}, MobileNet-v2\cite{sandler2018mobilenetv2} and Xception\cite{chollet2017xception} using pre-trained weights from 3.2 million ImageNet Data-set consisting of 1000 categories\cite{russakovsky2015imagenet} to compare with our achieved results (Table \ref{table3} and \ref{table4}), while we train Optic-Net from scratch with randomly initialized weights.

\begin{table}[bp]
\centering
\caption{Results on Srinivasan2014\cite{sri2014} Dataset}
\begin{tabular}{|c|c|c|c|}
\hline
Architectures & Test Accuracy & Sensitivity & Specificity \\ \hline
Lee et al. \cite{lee2017deep} & 87.63 & 84.63 & 91.54 \\ \hline
Awais et al. \cite{awais2017classification} & 93.00 & 87.00 & \textbf{100.00} \\ \hline
ResNet50-v1 \cite{he2016deep} & 94.92 & 94.92 & 97.46 \\ \hline
Karri et al. \cite{karri2017transfer} & 96.00 & -- & -- \\ \hline
MobileNet-v2 \cite{sandler2018mobilenetv2} & 97.46 & 97.46 &  98.73 \\ \hline
Xception \cite{chollet2017xception} & 99.36 & 99.36 & 99.68 \\ \hline
\textbf{OpticNet-71 [Ours]} & \textbf{100.00} & \textbf{100.00} & \textbf{100.00} \\ \hline

\end{tabular}
\label{table4}
\end{table}

\subsection{Hyper-parameter Tuning and Performance Evaluation}
The hyper-parameters while training OpticNet-47, OpticNet-63, OpticNet-71, MobileNet-v2 \cite{sandler2018mobilenetv2}, XceptionNet \cite{chollet2017xception}, ResNet50-v2 \cite{he2016identity}, ResNet50-v1 \cite{he2016deep} are as follows: batch size, b = 8; epochs = 30; learning rate, $\alpha^{lr} = 1e^{-4}$; step decay, $\gamma = 1e^{-1}$. We use adaptive learning rate and decrease it using $\alpha_{new}^{lr} = \alpha_{current}^{lr} \times \gamma$, if validation loss doesn't lower for six consecutive epochs. Moreover, we set the lowest learning rate to $\alpha_{min}^{lr} = 1e^{-8}$. Furthermore, We use Adam  optimizer with default parameters of $\beta_1^{adam} = 0.90$ and $\beta_2^{adam} = 0.99$ for all training schemes. We train OCT2017\cite{kermany2018identifying} data-set for 44 hours and Srinivasan2014\cite{sri2014} data-set for 2 hours on a 8 GB NVIDIA GTX 1070 GPU.

Inception-v3 models under-perform compared to both pre-trained models and OpticNet-71 as seen in Table \ref{table3}. OpticNet-71 takes 0.03 seconds to make prediction on an OCT image - which is real time and while accomplishing state-of-the-art results on OCT2017\cite{kermany2018identifying}, Srinivasan2014\cite{sri2014} data-set we also surpass human level prediction on OCT images as depicted in Table \ref{table3}. Human experts are real diagnosticians as reported in \cite{kermany2018identifying}. In \cite{kermany2018identifying}, there are 6 diagnosticians and the highest performing one is Human Expert 5 while the lowest performing one is Human Expert 2. To validate our CNN architecture's optimization strength we also train two smaller versions of OptcNet-71 on both dataests, which are OpticNet-47 ( [$N_1$ $N_2$ $N_3$ $N_4$] = [2 2 2 2] ) and OpticNet-63 ( [$N_1$ $N_2$ $N_3$ $N_4$] = [3 3 3 3] ). In Fig.~\ref{fig4} we unfold how all the variants of OpticNet outperforms the pre-trained CNNs on Srinivasan2014\cite{sri2014} data-set while OpticNet-71 outperforms all the pre-trained CNNs on OCT2017\cite{kermany2018identifying} data-set in terms of accuracy as well as performance-memory trade-off.

\subsection{Analysis of Proposed Residual Interpolated Block}
To understand how the Residual Interpolated Block works, we visualize features by passing a test image through our CNN model. Fig.~\ref{fig5}(a) illustrates some of the sharp signals propagated by Residual blocks while the interpolation reconstruction routine propagates a weak signal activation, yet the resulting signal space is both more sharp and fine grained compared to their Residual counterparts. Since the conv layers in the following stage activates the incoming signals first, we do not output a activated signal space from a stage. Instead we only activate the interpolation counterpart and then multiply with the last residual block's non-activated output space while adding the raw signals with the multiplied signal as well - which we consider as output from each stage as narrated in Fig.~\ref{fig5}(b). Furthermore, Fig.~\ref{fig5}(b) portrays how element-wise addition with the element-wise multiplication between signals helps the learning propagation of OpticNet-71. Fig.~\ref{fig5}(b) precisely depicts why this optimization chain is particularly significant, as a zero activation can cancel out a live signal channel from the residual counterpart ( $\tau(X_{l}) = \alpha(X_{l})+\beta(X_{l})\times(1+\alpha(X_{l}))$ ) while a dead signal channel can also cancel out a non-zero activation from the interpolation counterpart ( $\tau(X_{l}) = \beta(X_{l})+\alpha(X_{l})\times(1+\beta(X_{l}))$ ) - thus preventing all signals of a stage from dying and resulting in catastrophic optimization failure due to dead weights or gradient explosion.

\section{Conclusion}
In this work, we propose a novel convolutional neural network that potentially assists in demystifying abstraction from different retinal diseases and helps to identify them with human-level precision in real time. Moreover, we incorporate two novel architectural ideas to address the issue of gradient explosion and degradation. We hope to extend this work to conduct boundary segmentation of retinal layers, so that more subtle features and abnormalities can be detected autonomously with higher certainty which can be a potential tool for ophthalmologists around the world.
\section*{Acknowledgment}
We would like to thank \href{http://ccse.iub.edu.bd/}{``Center  for  Cognitive  Skill  Enhancement" Lab} for providing us with the technical support.

\bibliographystyle{IEEEtran}
\bibliography{reference}

\end{document}